\documentclass[11pt,a4paper]{article}
\pdfoutput=1

\usepackage{amsmath,amssymb}
\usepackage{epsfig,graphicx}
\usepackage{subfigure}
\usepackage{graphicx}
\usepackage{soul}
\usepackage{rotating}
\usepackage{cancel}
\usepackage{bm}
\usepackage{color}
\usepackage{comment}
\usepackage{cite}
\usepackage{psfrag}

\newcommand{\field}{\phi}

\renewcommand\[{\left[}

\newcommand{\exclude}[1]{}

\def\beq{\begin{equation}}
\def\eeq{\end{equation}}

\begin{document}
\numberwithin{equation}{section}
\title{{\huge{\bf{Foamy Dark Matter from Monodromies
}}}
\vspace{2.5cm} 
\Large{\textbf{
\vspace{0.5cm}}}}
\author{J\"urgen Berges, Aleksandr Chatrchyan\thanks{chatrchyan@thphys.uni-heidelberg.de}, Joerg Jaeckel\\[2ex]
\small{\em Institut f\"ur theoretische Physik, Universit\"at Heidelberg,} \\
\small{\em Philosophenweg 16, 69120 Heidelberg, Germany}\\[0.5ex]
}

\date{}
\maketitle

\begin{abstract}
\noindent
We investigate the dynamics of axion-like particle (ALP) dark matter where the field range is enlarged by a monodromy.
The monodromy potential allows sufficient production of dark matter also at larger couplings to the Standard Model particles. The potential typically features a number of ``wiggles'' that lead to a rapid growth of fluctuations. Using classical-statistical field theory simulations we go beyond the linear regime and treat the system in the non-linear and even non-perturbative regime. For sufficiently strong wiggles the initially homogeneous field is completely converted into fluctuations. The fluctuations correspond to dark matter particles with a non-vanishing velocity and we consider the corresponding restrictions from structure formation as well as the effects on today's dark matter density. Since all the dark matter is made up from these strong fluctuations, the dark matter density features large, $\mathcal{O}(1)$ fluctuations at scales $\lesssim 10^{6}\,{\rm km}\sqrt{{\rm eV}/m_a}$.
\end{abstract}

\newpage

\section{Introduction}
\label{sec:dm_intro}

Axion-like particles (ALPs) appear in quantum field theory as pseudo Nambu-Goldstone bosons (pNGBs) of spontaneously broken global symmetries\footnote{In the following we will use pNGB and ALP synonymously.}. 
PNGBs are a prototype of a particle where a high fundamental physics scale is linked to both a very weak interaction and a small mass, both being suppressed by the same high scale. 
An important example is the axion in quantum chromodynamics (QCD) solving the strong CP problem of QCD~\cite{Peccei:1977hh,Weinberg:1977ma,Wilczek:1977pj}. 
ALPs are also ubiquitous in string theory~\cite{Svrcek:2006yi,Arvanitaki:2009fg,Acharya:2010zx,Cicoli:2012sz} where they arise as Kaluza-Klein zero modes of antisymmetric tensor fields or as partners of moduli. 

Exact Goldstone bosons as well as their stringy counterparts enjoy a shift symmetry which protects them from perturbative corrections to their mass.
Explicit symmetry breaking, e.g. due to non-perturbative effects, reduces the continuous shift symmetry of ALPs down to a discrete one. For an ALP described by a scalar field $\phi$, the shift
\begin{equation}
\field \rightarrow \field + 2\pi f
\end{equation}
then leaves the field's potential invariant. The potential of ALPs is therefore periodic, and can typically be parametrized as 
\begin{equation}
U(\field)=\Lambda^4 [1-\cos(\field/f)].
\end{equation}
Here, the decay constant $f$ is the scale of the spontaneous symmetry breaking, or the string scale for string axions. $\Lambda$ is the scale of the explicit symmetry breaking effects.
This potential corresponds to a mass,
\begin{equation}
m_a^2=\frac{\Lambda^4}{f^2},
\end{equation}
which is suppressed by the large axion decay constant.
In a similar manner, the shift symmetry ensures that all couplings of ALPs are suppressed by powers of the decay constant.

ALPs can also play an important role in cosmology (for a review see, e.g.~\cite{Marsh:2015xka}). 
If produced via the misalignment mechanism~\cite{Preskill:1982cy,Abbott:1982af,Dine:1982ah,Arias:2012az}, they are good candidates for cold dark matter (CDM). 
In the simplest case of this mechanism, the axion-like field is present already during inflation. After inflation it obtains some value $\phi_1$, practically homogeneous throughout the observable universe. 
The field performs coherent oscillations around the minimum of the potential, once the Hubble parameter drops below its mass.
The amplitude of oscillations decays due to expansion and very soon the oscillations take place in the almost quadratic region of the potential. 
As a result, the field behaves exactly as ordinary matter and, in particular, dilutes according to $\rho \propto a^{-3}$. The energy density of ALPs today can be expressed in terms of their mass $m_a$ and the misalignment field value $\phi_1$ as~\cite{Arias:2012az}
\beq
\label{eq:dmdensitytd}
\rho_{\mathrm{today}} \approx 0.17 \,\frac{\mathrm{keV}}{\mathrm{cm}^3} \sqrt{\frac{m_a}{\mathrm{eV}}} \Bigl(\frac{\phi_1}{10^{11}\mathrm{GeV}}\Bigr)^2.
\eeq
This quantity can be compared with the observed dark matter (DM) density today, $\rho_{ \mathrm{CDM} } \approx 1.27\, \mathrm{keV/ cm}^3$ \cite{Aghanim:2018eyx}. 

\begin{figure}[!t]
\centering
\includegraphics[height=0.49\textheight]{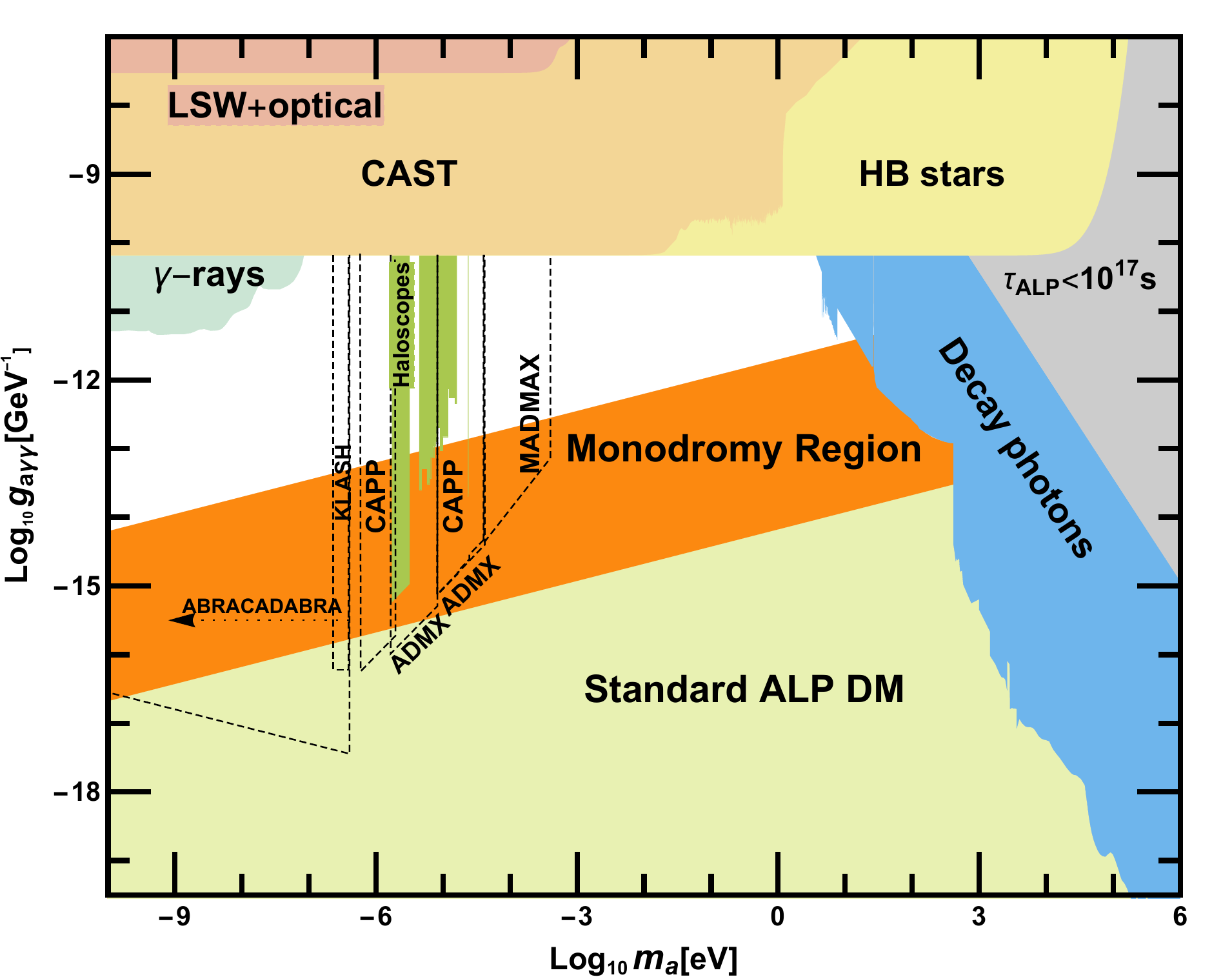}
\caption{Allowed values of coupling and mass for ALPs. Upcoming dark matter experiments are indicated by the dashed black lines. Compilation adapted from~\cite{Redondo:2008en,Jaeckel:2010ni,Arias:2012az,Irastorza:2018dyq}. The region labeled ``Monodromy'' is allowed if the field range of the ALP is increased~\cite{Jaeckel:2016qjp} up to $\sim10^3\,f$. }
\label{allowed}
\end{figure}

Their periodic potential limits their field range to $|\phi_1| < \pi f$. Therefore, requiring that ALPs are most or all of dark matter puts an upper limit on the couplings that are $\sim 1/f$~\cite{Arias:2012az}. 
For the example of an ALP-photon coupling,
\begin{equation}
{\mathcal{L}}_{\mathrm{int}} = -\frac{1}{4}g_{a\gamma\gamma}\phi F^{\mu\nu}\tilde{F}_{\mu\nu},\qquad g_{a\gamma\gamma}=c_{a\gamma}\frac{\alpha}{4\pi f},
\end{equation}
this is indicated in Fig.~\ref{allowed} as the yellow "Standard ALP DM"~\cite{Arias:2012az} region, where we have chosen the model dependent ${\mathcal{O}}(1)$ constant $c_{a\gamma}$ to be equal to unity.

Upcoming dark matter experiments (indicated by dashed lines in Fig.~\ref{allowed}) will mostly reach only couplings larger than those expected for Standard ALP DM.
It is therefore interesting to see what models would allow ALPs to be the dark matter while having such larger couplings.
One possibility is a temperature dependent mass as predicted for the QCD axion, the latter being within the reach of most of the indicated dark matter experiments.
This has already been discussed in~\cite{Arias:2012az} and can lead to a significant increase in the viable dark matter region. However, for light particles this likely requires a non-trivial sector responsible for the mass generation that has a significant temperature also at relatively late times. Other possibilities include non-canonical kinetic terms~\cite{Alvarez:2017kar}, gravitational couplings to the Ricci scalar $R$~\cite{Alonso-Alvarez:2018tus}, mixing of multiple ALPs~\cite{Ho:2018qur} and other modifications of the minimal setup~\cite{Daido:2017wwb,Daido:2017tbr,Daido:2018dmu}.

An interesting alternative is the possibility of having an explicitly broken discrete shift symmetry. 
Such ALPs are said to exhibit a monodromy and have been realized for string axions~\cite{McAllister:2008hb, Marchesano:2014mla} and applied to inflationary model building~\cite{McAllister:2008hb, Silverstein:2008sg}.
For dark matter, ALPs with a monodromy have been discussed in~\cite{Jaeckel:2016qjp}. 
For our case at hand the important effect is that the constraint $\phi_1 < \pi f$ is absent and the potential is not bounded from above, but can take larger values.
This allows one to bypass the limitation and realize ALP DM also at larger couplings~\cite{Jaeckel:2016qjp} into the orange region shown in Fig.~\ref{allowed}.
This is the scenario we want to investigate in more detail in this paper.

An  important feature of the potential for an ALP with a monodromy is that it consists of a periodic term plus an extra quadratic term as shown in Fig.~\ref{potwiggles}.
As already noted in~\cite{Jaeckel:2016qjp} the evolution of the field in such a potential leads to a rapid growth of fluctuations, even if the field is initially homogeneous as expected if the field is already present during inflation. 
This is where this type of monodromy DM differs decisively from ordinary ALP DM, where the field and the corresponding density remains essentially homogeneous until structure formation sets in.
This work is dedicated to studying the growth of fluctuations and determining phenomenological consequences from it.
Here we go far beyond the initial analysis of~\cite{Jaeckel:2016qjp} which was based on linearized equations for the fluctuations and therefore could only study the initial growth. Instead we use classical-statistical field theory simulations in an expanding space-time to proceed into the non-linear regime.

We find three important features.
\begin{itemize}
\item{} For a wide range of parameters there is very rapid growth of fluctuations. This corresponds to the production of relativistic particles.
However, for most masses ($m\gtrsim 10^{-15}\,{\rm eV}$) these particles have cooled sufficiently by matter-radiation equality such that they remain fully viable as cold dark matter. 
\item{} The reduction in energy density due to this relativistic intermediate phase is relatively small.
\item{} The density fluctuations can easily be of order one. However, the typical scales are much smaller than those expected for the formation of axion miniclusters~\cite{Hogan:1988mp} (cf.~also e.g.~\cite{Kolb:1994fi,Enander:2017ogx,Fairbairn:2017sil}). At these small scales the localization of the low mass requires them to have a non-vanishing momentum and the resulting pressure prevents the collapse. We therefore do not expect miniclusters with huge over-densities.
Instead we have a rather inhomogeneous dark matter with ${\mathcal{O}}(1)$ fluctuations at scales $\sim (10^{3}-10^{6})\,{\rm km}\sqrt{{\rm eV}/m_a}$. The evolution of these fluctuations during structure formation and possible consequences need further study.
\end{itemize}

The remainder of this work is organized as follows. In Section~\ref{ssec:misalignment} we describe the considered model of monodromy ALPs and the scenario of the misalignment mechanism.
In Section~\ref{ssec:dm_growth} we study the dynamics of vacuum realignment and the growth of fluctuations in the presence of monodromy. 
We determine the impact of the fluctuations on the equation of state and on the dark matter energy density today in Section~\ref{ssec:eos}. 
The power spectrum and the evolution of the over-densities are discussed in Section~\ref{ssec:minicluster}. We conclude in Section~\ref{ssec:conclusion}. We use $\hbar=c=1$ and consider a flat background Friedmann-Robertson-Walker (FRW) metric $ds^2=dt^2-a^2(t)d\mathbf{x}^2$ with scale factor $a(t)$.

\section{ALPs with a monodromy and the misalignment mechanism}
\label{ssec:misalignment}
Our starting point is a monodromy potential where the discrete shift symmetry of ALPs is broken by a quadratic monomial. The classical potential can be written as
\begin{equation}
\label{eq:potentialform}
U(\field)=\frac{1}{2}m^2\field^2 + \Lambda^4 \Bigl( 1-\cos { \frac{\field}{f} } \Bigr).
\end{equation}
This corresponds to the so-called massive Sine-Gordon model. In principle, an arbitrary phase parameter can be included in the argument of the cosine. We set it to zero for simplicity, which leads to a $\mathbb{Z}_2$-symmetric potential with a global minimum at $\field=0$. 

\begin{figure}[!t]
\centering
\label{potwiggles}
\includegraphics[width=0.58\textwidth]{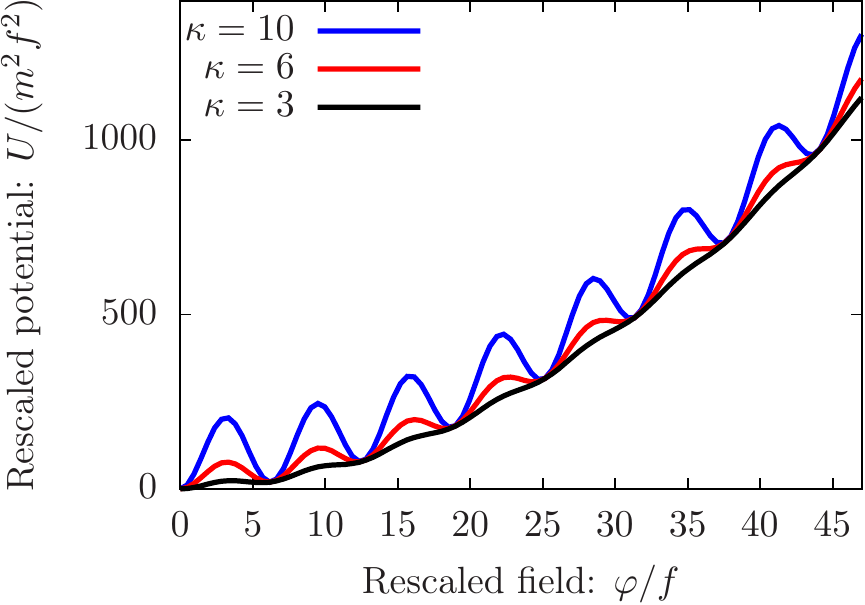}
\caption{The shape of the rescaled potential $x^2/2 + \kappa^2(1-\cos {x} )$ for several values of $\kappa$.}
\end{figure}

The relative strength of the quadratic term compared to the periodic one can be quantified by the value of the dimensionless parameter $\kappa$, defined as 
\beq
\kappa^2 = \frac{\Lambda^4}{m^2f^2}.
\eeq
This parameter determines the strength of the wiggles as can be seen from Fig.~\ref{potwiggles}. The potential develops an increasing number of local minima as $\kappa$ grows beyond $\kappa\sim 1$.

Particle excitations around the global minimum have a mass that is given by,
\beq
m_a^2 =\frac{\delta^2 U}{\delta \phi^2} \Big|_{\phi=0}= m^2(1+\kappa^2). 
\eeq
Due to the cosine potential the monodromy ALP also features a whole sequence of self-interactions. For example we obtain for the four- and six-field interactions,
\begin{equation}
\lambda_{4}=\frac{\kappa^2}{1+\kappa^2}\frac{m^{2}_{a}}{f^2},\qquad \lambda_{6}=\frac{\kappa^2}{1+\kappa^2}\frac{m^{2}_{a}}{f^4}.
\end{equation}
Since we have $m_{a}\ll f$ all interactions are very weak. In particular we have $\lambda_{4}\ll 1$.

\bigskip
In an expanding FRW universe the classical equations of motion in comoving coordinates read,
\beq
\label{FRWeq}
\ddot \field + 3 H \dot \field - \frac{\Delta \field}{a^2}  +\frac{\delta U}{\delta \field} =0,
\eeq
where $a(t)$ is the scale factor and $H=\dot a/a$ is the Hubble parameter. 

In the following we consider the scenario where the axion-like field is already present during inflation~\cite{Arias:2012az}. 
This implies $f \gg H_I$, where $H_I$ is the inflationary Hubble parameter.

After inflation the initial condition is then given by,
\begin{equation}
\field(x)=\field_{1}={\rm const},
\end{equation}
which is constant throughout the entire observable universe. The subsequent evolution of $\phi(t)$ is approximately governed by \eqref{FRWeq}, where the gradient term can be neglected due to homogeneity.
The field is initially over-damped, $\partial_t \field=0$, and only starts to oscillate once the Hubble friction term becomes comparable to the potential term in the equations of motion. 
This happens approximately when $H=H_{\textrm{osc}} = m_a/3$, although it can be delayed depending on the value of $\phi_1$ due to the wiggly structure of the potential.
For later use, we denote the scale factor at this time by $a_{\textrm{osc}}$.

ALPs that are expected to behave as cold dark matter should start to oscillate when the universe is still radiation-dominated, prior to the epoch of matter-radiation equality at $H_{\mathrm{eq}}=2.3 \times 10^{-28} \mathrm{eV}$. Otherwise, the axion-like field would behave as dark energy during matter-radiation equality. However, as we assume the mass to be constant this does not pose any serious additional constraint since $m_{a}\gtrsim  10^{-22}\,{\rm eV}$ is required by constraints from structure formation~\cite{Marsh:2015xka}.

Solving the equations of motion for the approximation of a quadratic potential with a constant mass yields Eq.~(\ref{eq:dmdensitytd}).
Deviations from this due to the non-quadratic form of the monodromy potential have been discussed in~\cite{Jaeckel:2016qjp}. In Section~\ref{ssec:energy_density} we will study the deviations caused by the growth of fluctuations.

\bigskip
So far we have discussed the initial conditions and the evolution of the homogeneous field. However, as already mentioned due  to their growth fluctuations are highly relevant in the case of sufficiently wiggly monodromy potentials.
In general all fields feature unavoidable quantum fluctuations. In particular, in the early universe
inflation imprints Hubble sized fluctuations in the axion-like field.
More precisely, as $m_a \ll H_I$, we can treat the ALP as massless during inflation. Quantum fluctuations of a massless scalar field in de-Sitter spacetime are known to have an approximately scale-independent field power spectrum (cf. e.g.~\cite{Baumann:2009ds}),
\beq
\Delta_{\field}= \Bigl(\frac{H_I}{2\pi}\Bigr)^2.
\eeq
Here, the power spectrum is defined as\footnote{We denote the quantum operators as hatted fields. The fields without hat correspond to the
expectation values $\phi=\langle\hat{\phi}\rangle$. In the classical-statistical field theory simulations the quantum expectation value is obtained from averaging over the statistical samples with the initial correlation functions fixed by the quantum theory~\cite{Berges:2015kfa}.}
\beq
\label{eq:defineFP}
\frac{1}{2} \langle \{ \hat \phi_{\mathbf{p}}(t), \hat \phi_{\mathbf{p'}}(t')  \}\rangle|_{t'=t} =(2\pi)^3 \delta^{(3)}(\mathbf{p}+\mathbf{p'})F(t,t',\mathbf{p})|_{t'=t} = (2\pi)^3 \delta^{(3)}(\mathbf{p}+\mathbf{p'}) \frac{2\pi^2}{\mathrm{p}^3} \Delta_{\field}(t, \mathbf{p}).
\eeq
In this equation $\hat{\phi}_{\mathbf{p}}(t)$ are the field modes in Fourier space and $F(t, t', \mathbf{p})$ is the statistical propagator. After inflation, all momentum modes of interest are super-horizon and remain frozen, $\partial_t F(t,t',\mathbf{p})=0$, until they re-enter the horizon.

Let us already note at this point that for our purposes (cf.~\cite{chatrchyaninprep}) all stages of the evolution of interest to us can be described to excellent precision by treating all the quantum fields as classical-statistical fields. We can therefore think of the Fourier modes of the field simply as the classical fluctuation modes and solve the classical equation of motion, Eq.~\eqref{FRWeq}, multiple times. Power spectra and occupation numbers can then be directly obtained by evaluating the average in Eq.~\eqref{eq:defineFP} for a statistical sampling of initial conditions.

\section{Growth of fluctuations}
\label{ssec:dm_growth}
Having set the stage we now turn to the behavior of the fluctuations which forms the core of this work.

\subsection{Linear regime}
\subsubsection*{Treatment of fluctuations in the linear regime}
As already noted we are in a situation where $H_{I}\ll f$. Therefore the fluctuations $\sim H_{I}$ from inflation given in Eq.~\eqref{eq:defineFP} are much smaller than
the scale $\sim f$ at which non-linearities set in.
Initially, the dynamics can therefore be studied with the help of linearized equations of motion~\cite{Jaeckel:2016qjp}. This is conveniently done in terms of conformal variables,
\beq
\hat \phi_{\mathrm{c}} = a \hat \phi, \: \: \: \: \: \: \: \: \: \: \: d\tau = dt/a.
\eeq
The linearized equations for the momentum modes of the statistical propagator $F_{\mathrm{c}}(\tau,\tau',\mathbf{p}) =a(t)a(t')F(t,t',\mathbf{p})$ then take the form~\cite{Tranberg:2008tg}
\beq
\label{linearizede}
\Bigl[\partial^2_{\tau}+ \mathbf{p}^2+a^2 m^2\Bigl(1+\kappa^2 \cos{\frac{\phi}{f}}\Bigr) - \frac{1}{a} \frac{d^2a}{d\tau^2}\Bigr] F_{\mathrm{c}}(\tau, \tau', \mathbf{p}) = 0,
\eeq
where $\mathrm{p}$ is the comoving momentum, related to the physical one by $\mathrm{p}_{\mathrm{phys}}=\mathrm{p}/a$. The above expression differs from the Minkowski analogue~\cite{chatrchyaninprep} only by the $a$-dependent modification of the effective mass. To study the growth of the fluctuations we have solved (\ref{linearizede}) numerically, for a wide range of momenta, together with the classical\footnote{The back-reaction of the fluctuations on the background field is negligible at this stage} evolution equation for the background field $\phi_c(\tau)=a(t)\phi(t)$,
\beq
\partial^2_{\tau} \phi_c +\Bigl( a^2 m^2 -   \frac{1}{a} \frac{d^2a}{d\tau^2}\Bigr) \phi_c +a^3 \frac{\Lambda^4}{f} \sin \Bigl( \frac{\phi_c}{a f}\Bigr)=0.
\label{conformeq}
\eeq
For the scale factor in a radiation-dominated universe we have taken
\beq
\label{eq:scale_factor}
a = \sqrt{1+2 H_0 (t-t_0)} = 1+H_0 (\tau-\tau_0),
\eeq
which in addition to $a(t) \sim t^{1/2}$ fixes $a(t_0)=1$ and $H(t_0)=H_0$. Here the $0$-index corresponds to the initial time of the simulation and $H_{0}$ is the value of the Hubble parameter at that time. From (\ref{eq:scale_factor}) one can check that the terms containing second derivatives of the scale factor in (\ref{linearizede}) and (\ref{conformeq}) vanish for a radiation-dominated expansion. To properly account for the horizon re-entry of all modes of interest, we have chosen $H_0$ to be of the same order as the largest momentum of the modes. For $\phi$ and $F$ we have employed the set of initial conditions described in the previous section.

In terms of the conformal variables the occupation numbers can be defined completely analogous to the case of Minkowski spacetime~\cite{Tranberg:2008tg}, 
\beq
\label{occupe}
n(t(\tau),\mathbf{p})+1/2 = \sqrt{F_{\mathrm{c}}(\tau,\tau',\mathbf{p}) \partial_{\tau} \partial_{\tau'} F_{\mathrm{c}}(\tau,\tau',\mathbf{p})} |_{\tau'=\tau}.
\eeq
Defined in this way these are occupation numbers per comoving volume and, therefore, for sub-horizon modes, they do not change due to expansion. This is different for  super-horizon modes. In this case the occupation number grows with time, which is a consequence of the Hawking radiation emitted from the horizon~\cite{Gibbons:1977mu}.

When solving numerically the evolution equations, all dimensionful quantities are expressed in units of $m$, and the decay constant is rescaled out from the equations after the transformation $\phi \rightarrow \phi/f$. As a result the dynamics depends only on the values of three dimensionless parameters: $\kappa$, $\phi_1/f$ and $H_I/f$, with the latter determining the initial strength of the fluctuations. 

\subsubsection*{Growth of fluctuations in the linear regime}
Having set up the technique to deal with the fluctuation in the linear regime let us now return to the physics.
For small values of $\kappa\ll 1$ the evolution proceeds essentially as for a quadratic potential, i.e.~a free field. The dynamics is dominated by the behavior of the homogeneous field.
We are, however, more interested in the case where the potential is non-trivial. As a benchmark case we take,
\begin{equation}
\kappa=10.
\end{equation}
The initial condition for the field and its fluctuations can be specified by fixing the initial field value as well as the Hubble scale\footnote{We note that large-scale isocurvature fluctuations from inflation are not problematic for $H_{I}\lesssim 10^{6}\,{\rm GeV}$ (cf., e.g.~\cite{Arias:2012az,Alvarez:2017kar}).},
\begin{equation}
\field_{1}/f=1000,\qquad H_{I}/f=10^{-10}.
\end{equation}

\begin{figure}[!t]
\centering
\includegraphics[width=0.55\textwidth]{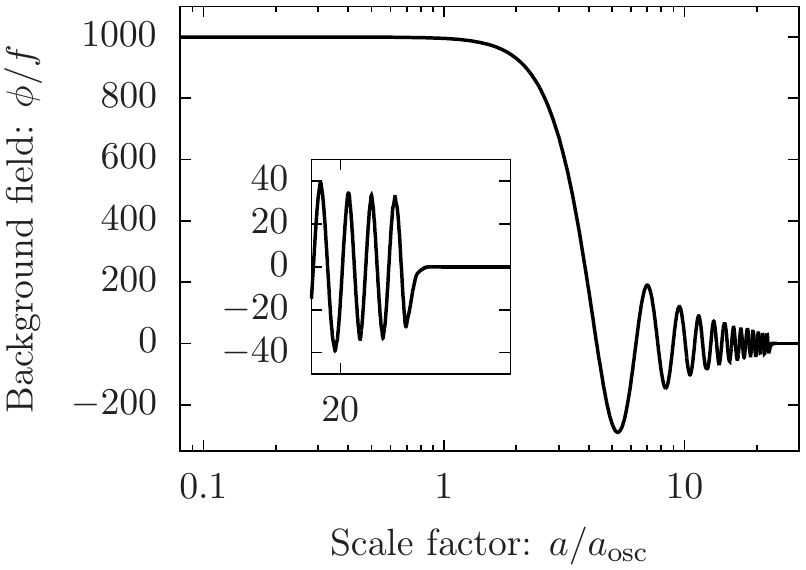}
\caption{The evolution of the background field $\phi(t)/f$ as a function of the scale factor $a(t)/a_{\mathrm{osc}}$, demonstrating the transition from the "over-damped" to the oscillatory phase, as well as the collapse of the background field in the non-perturbative regime with strong fluctuations. The collapse, which occurs at $a \approx 22 a_{\mathrm{osc}}$, is shown with a better resolution in the inset. We employ $\phi_1/f=1000$, $\kappa=10$ and $H_I/f=10^{-10}$. }
\label{back21}
\end{figure}

The evolution of the background field in the linear regime is straightforward. It starts to oscillate around $a_{\mathrm{osc}}$ and is then damped by the Hubble expansion as can be seen in Fig.~\ref{back21}, where the linear regime extends up to $a < 19\,a_{\mathrm{osc}}$. 

The behavior of the fluctuations  is more interesting.
Once the background field starts to oscillate, fluctuations are subject to a parametric resonance instability~\cite{Berges:2002cz}\footnote{We do not consider the production of Standard Model particles arising due to their coupling to the ALP field. For sufficiently weak couplings this should be a small effect. For example, for the case of a photon coupling the relevant parameter for the growth of fluctuations is $g_{a\gamma\gamma}\phi \sim \alpha\phi/(2\pi f)$, which is $\lesssim 1$ in  our case. Significant growth is  expected only
if it is $\gtrsim 10$~\cite{Agrawal:2017eqm,Agrawal:2018vin,Co:2018lka, Bastero-Gil:2018uel}.}
The resonance in the absence of expansion was studied in~\cite{Jaeckel:2016qjp} and was shown to exhibit a number of instability bands. 
The same holds in the presence of expansion. However, there are two essential modifications.
The first is that the amplitude of the oscillations of $\phi$ decreases with time due to expansion, thus the effective mass of the fluctuations in \eqref{linearizede} is not strictly periodic. Second, in terms of conformal variables, expansion causes the increase of characteristic mass scales in \eqref{linearizede} proportional to the scale factor, reflecting the red-shift of physical momenta. 

\begin{figure}[!t]
\centering
\includegraphics[width=0.701\textwidth]{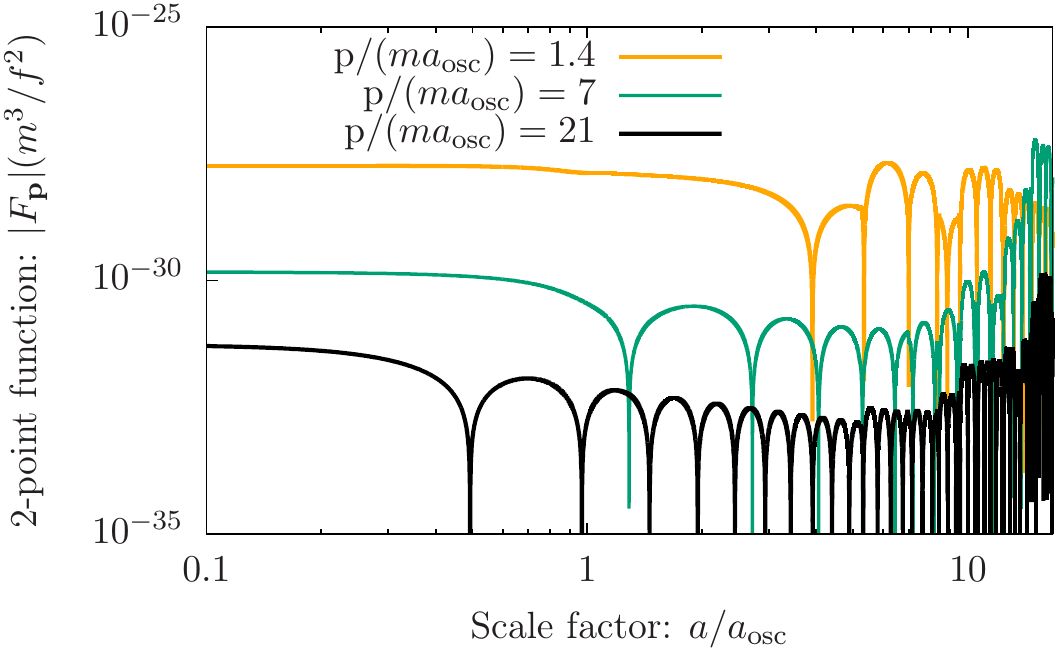}
\caption{The linear evolution of absolute values of several momentum modes of the statistical propagator $|F(t,t_0,\mathbf{p})|(m^3/f^2)$ as a function of the scale factor $a(t)/a_{\mathrm{osc}}$, demonstrating the horizon re-entry of the modes, as well as their subsequent enhancement due to instabilities. We employ $\phi_1/f=1000$, $\kappa=10$ and $H_I/f=10^{-10}$.}
\label{F21}
\end{figure}

In Fig.~\ref{F21} we plot the linear evolution of several momentum modes of the two-point function $F(t,t_0,\mathbf{p})$, defined in \eqref{eq:defineFP}. After the horizon re-entry the modes start to oscillate with an amplitude decreasing with expansion. As expected, modes with higher momenta re-enter the horizon earlier. 
Once a mode enters an instability band it starts to grow until the continuing red-shifting moves it out of the instability band again. Such an intermittent growth can be observed in Fig.~\ref{F21}. The intermittent nature is most apparent in the high momentum mode shown in black.
The wider the instability band the more time is spend in growing the mode.
Therefore the high-momentum narrow instability bands are less efficient compared to the lowest-momentum bands, where the dominant growth occurs.

\subsection{Non-linear regime}
The linear approximation becomes inaccurate once fluctuations become sufficiently large. The condition for the first non-linear corrections becoming important can be determined in an analogous way to the case of Minkowski spacetime~\cite{chatrchyaninprep}, with the difference that initially $F \sim \mathcal{O}(H_I^2)$. This implies that secondary instabilities~\cite{Berges:2002cz} set in when parametrically $F \sim \mathcal{O}(H_I f)$ for the strongest resonant modes. They correspond to re-scattering of the produced fluctuations and result in a broader and smoother spectrum in momentum space. The dynamics becomes non-perturbative if $F \sim \mathcal{O}(f^2)$. In this stage the background field transfers its energy to the fluctuations until it collapses due to the presence of attractive self-interactions~\cite{Berges:2017ldx}. 

\subsubsection*{Simulating the non-linear regime}

 To study the dynamics in the non-linear regime we have performed classical-statistical lattice simulations in an expanding FRW universe. We have developed a C++ program, similar to LATTICEEASY~\cite{Felder:2000hq}. It solves the classical equations of motion, using conformal variables, in terms of which expansion manifests itself only via the time-dependence of the mass scales. Red-shifting restricts the lattice simulations to not too late times, such that the mass scale is still within the resolution of the lattice.  Our program first solves the linearized equations without taking into account their back-reaction on the background field. Before the fluctuations become non-linear, i.e.~secondary instabilities set in, we switch to the lattice simulation. The field configuration obtained from the linear evolution is used as an initial condition for the lattice simulation. (See~\cite{Berges:2015kfa} for more details on how to realize such initial conditions in a classical-statistical simulation.) This allows us to partially avoid running out of lattice points due to the rapid red-shifting at early times\footnote{Note that expansion is the most rapid at early times.} and to evolve the field longer.  We have used cubic lattices with up to $1024^3$ points, periodic boundary conditions and a fixed comoving volume.

As for the linear evolution, the dynamics in terms of rescaled dimensionless quantities depends only on $\kappa$, $\phi_1/f$ and $H_I/f$.

\subsubsection*{Evolution in the non-linear regime}
We now consider the behavior in the non-linear regime.
The axion-like field gets diluted with time due to expansion. After some time it explores the approximately quadratic region of the potential near its minimum. Self-interactions become extremely weak at this stage and the evolution of the occupation numbers slows down. For small values of $\phi_1/f$, such ``freezing'' happens already after a couple of oscillations of the background field, which allows to ignore the growth of the fluctuations. This is the case for standard ALPs, for which the constraint $\phi_1/f < \pi$ holds.  In the presence of a monodromy the value of $\phi_1/f$ can be larger and, therefore, the occupation numbers can become large before they \lq\lq{freeze}\rq\rq.

Looking at the background field we observe a rapid drop of the amplitude to zero at $a/a_{\mathrm{osc}} \approx 23$. This arises when the fluctuations become sufficiently large and their growth rapidly extracts the energy from the background field.

\begin{figure}[!t]
\centering
\includegraphics[width=0.999\textwidth]{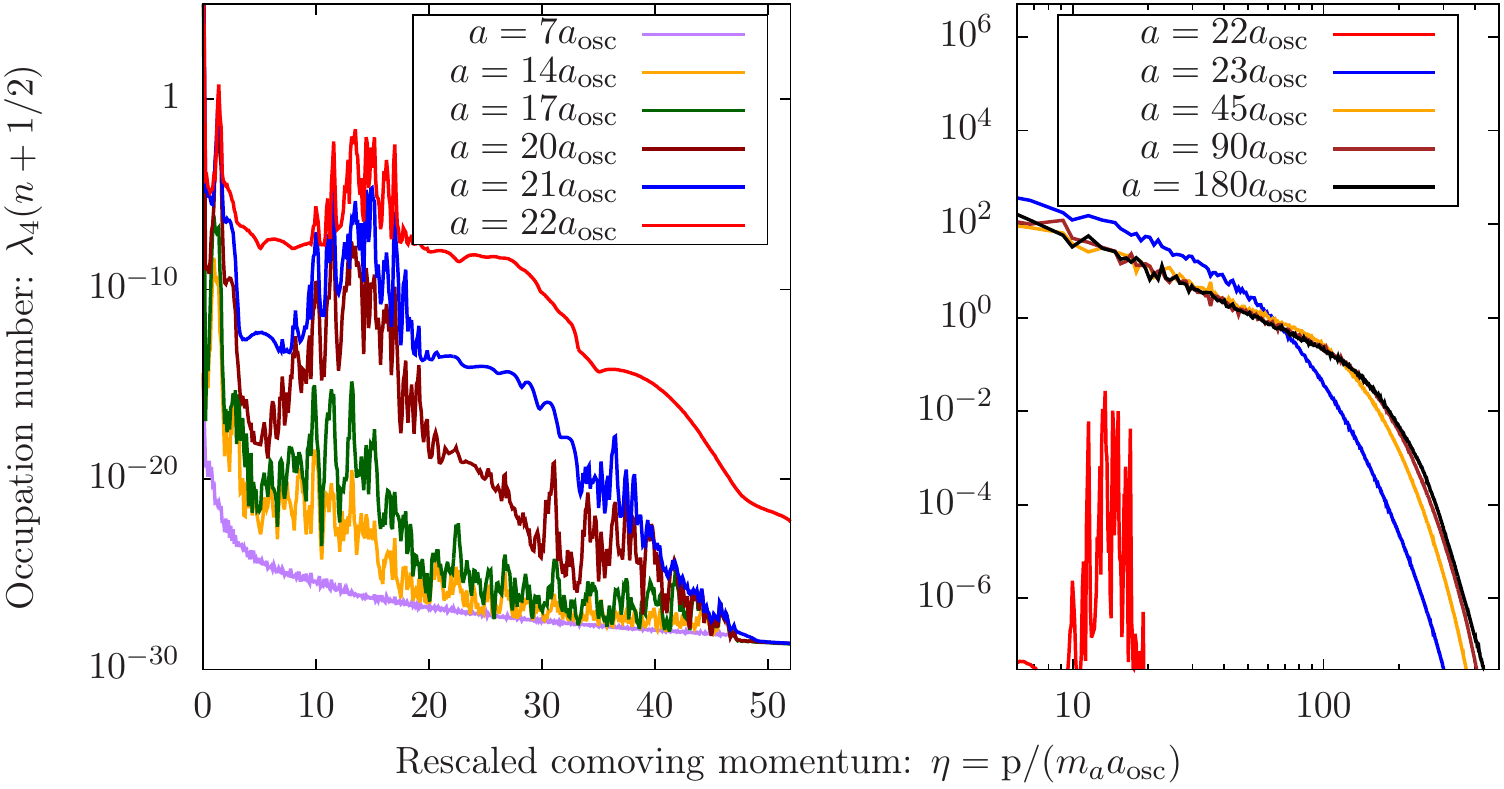}
\caption{Several snapshots of occupation numbers at different scale factors. We employ $\phi_1/f=1000$, $\kappa=10$ and $H_I/f=10^{-10}$.}
\label{spectrum}
\end{figure}

The evolution of the fluctuations can be best understood in terms of the occupation numbers~\eqref{occupe}. The occupation numbers at different times are shown in Fig.~\ref{spectrum}. At first we see the expected growth of fluctuations. We note that the broad bands in the left plot at intermediate times, $19<a/a_{\rm osc}<22$, are a result of secondary instabilities~\cite{chatrchyaninprep}. Once the background field has collapsed we observe a smooth cascade towards higher momenta, for $a/a_{\rm osc}>23$. This cascade is similar to~\cite{Micha:2004bv}. However, at late time the present cascade slows down extremely due to the dilution from the expansion.
When the density decreases, interactions slow down the evolution and the occupation numbers effectively freeze out and only continue to red-shift.
In terms of comoving momenta this simply means that the distribution of the occupation numbers remains constant as can be seen by comparing the late-time brown and black curves.

\section{Structure formation, the equation of state and today's energy density}
\label{ssec:eos}
In the previous section we have seen that the evolution of the field through the wiggly regions causes a massive growth of fluctuations and, in particular, of fluctuations with rather high momenta. As can be seen from Fig.~\ref{spectrum} comoving momenta up to $\mathrm{few} \times 100\, m_{a}$ are populated. Taking the expansion into account the physical momenta are still up to $10 \,m_{a}$. In the particle picture this means that we are dealing with quite relativistic particles. 
This raises two questions: Will the particles be sufficiently non-relativistic to allow for successful structure formation, and how much of the gain in the energy density 
is lost due to the faster dilution of the energy density for relativistic species?
In this section we will address these questions.

\subsection{Sufficiently cool monodromy dark matter}
We start with the question about structure formation. Shortly after oscillations of the ALP field begin we have significant production of quite relativistic particles.
As we will see below, and as we would expect, this leads to a phase where we have a relativistic equation of state.
However, this immediately also raises the question whether these particles are too fast, i.e. hot, to allow for successful structure formation.

To obtain the constraints arising from this we estimate the typical velocity of our dark matter particles at matter-radiation equality and compare them to those of (allowed) warm dark matter.

As we can see from Fig.~\ref{spectrum}, the evolution of the ALP occupation numbers expressed in comoving momenta freezes at around $a/a_{\rm osc}\sim 100$.
This is due to the drop in interaction rates resulting from the dilution by the Hubble expansion. From then on we can take the spectrum to be essentially frozen.

The highest significantly occupied comoving momentum is about
\beq
\frac{\mathrm{p}_{\rm max}}{a_{\rm osc}}=\frac{a \mathrm{p}^{\rm phys}_{\rm max} }{a_{\rm osc}}\sim \mathrm{few} \times 100\,m_{a} .
\eeq

Neglecting changes in the number of degrees of freedom the Hubble scale during the radiation dominated epoch behaves as,
\begin{equation}
\label{scale_factor_Hubble}
H\sim \frac{T^{2}}{M_{\rm P}}\sim \frac{1}{a^2}.
\end{equation}
We can therefore estimate\footnote{We use $H_{\rm eq}=2.3\times 10^{-28}\,{\rm eV}$.},
\beq
\frac{a_{\rm osc}}{a_{\rm eq}}\sim\sqrt{\frac{H_{\rm eq}}{H_{\rm osc}}}=\sqrt{\frac{3 H_{\rm eq}}{m_a}}\sim 10^{-6} \left(\frac{10^{-15}\,{\rm eV}}{m_{a}}\right)^{1/2}.
\eeq
Using this  we find for the maximum velocity at matter-radiation equality,
\beq
v_{\rm eq, max}=\frac{\mathrm{p}_{\rm max}}{m_a a_{\rm eq}}\sim 10^{-3}\left(\frac{10^{-15}\,{\rm eV}}{m_{a}}\right)^{1/2}.
\eeq

Let us now compare this to the velocity of typical warm dark matter candidates at matter-radiation equality. Typical warm dark matter candidates have masses in the $\mathrm{keV}$ range (cf., e.g.~\cite{Viel:2005qj}). Assuming relativistic decoupling and again neglecting the changes in the number of degrees of freedom we can estimate,
\begin{equation}
v_{\rm warm\,\,DM,\,\,eq}\sim 10^{-3}\qquad{\rm for} \qquad m_{\rm warm\,\,DM}\sim{\rm few}\times{\rm keV}.
\end{equation}

Therefore we require
\beq
m_{a}\gtrsim 10^{-15}\,{\rm eV}\qquad{\rm for}\qquad \kappa\gtrsim 1
\eeq
in order to have successful structure formation for monodromy ALPs with strong wiggles, i.e. $\kappa\gtrsim 1$. We note that this limit does not apply if $\field_{1}/f \lesssim 10$, since in that case the production of the fluctuations is negligible.
The precise number depends somewhat on $\kappa$ as well as the initial field amplitude $\field_{1}/f$.

\subsection{Equation of state}
Even if velocities are sufficiently small to allow for structure formation to proceed, there is still an important effect of the growth of fluctuations and the production of relativistic particles: the equation of state is changed for some time after oscillations begin.
This is what we will study in this subsection.

The equation of state parameter is defined as 
\begin{equation}
w = p / \rho\,,
\end{equation}
where $p$ and $\rho$ are the expectation values of the pressure and the energy density, respectively. They are given by
\beq
\rho= \langle \frac{1}{2}\dot \field ^2 + \frac{1}{2} (\nabla \field )^2 + U(\field) \rangle,\: \: \: \: \: \: \: \: \: p=\langle \frac{1}{2}\dot \field ^2 - \frac{1}{6} (\nabla \field )^2 - U(\field) \rangle.
\eeq

Our baseline expectations are
that a constant homogeneous background field leads to $w \approx -1$. For a homogeneous background field oscillating around $\field=0$, $w$ oscillates between $1$ and $-1$. In the case of a purely quadratic potential these oscillations are such that the average over a few oscillations is $\bar w=0$, corresponding to a system dominated by non-relativistic fluctuations (matter). If ultra-relativistic fluctuations are dominant (radiation), $w \approx 1/3$. Taking all of this into account, we expect that:
\begin{itemize}
\item At early times $w=-1$.
\item After the background field starts to oscillate, $w$ oscillates between $-1$ and $1$.
\item As the background oscillation amplitude decreases, the field spends more time in the concave regions of the potential. In this regions potential energy is dominant over kinetic energy, which shifts $\bar w$ towards negative values (see also \cite{Jaeckel:2016qjp}).
\item The growth of fluctuations with large momenta shifts $\bar w$ in the opposite direction, towards $w=1/3$.
\item After the collapse of the background field, $w$ is between 0 and $1/3$.
\item Because of the redshift of momenta, $w \rightarrow 0$ at late times.
\item If the field settles in a false minimum and obtains an expectation value, the corresponding vacuum energy does not dilute with time and, therefore, will dominate the energy content after some time, leading to $w \rightarrow -1$. For our purposes this dark energy contribution must be subtracted from the total energy density.
\end{itemize}

\begin{figure}[!t]
\centering
\includegraphics[width=0.55\textwidth]{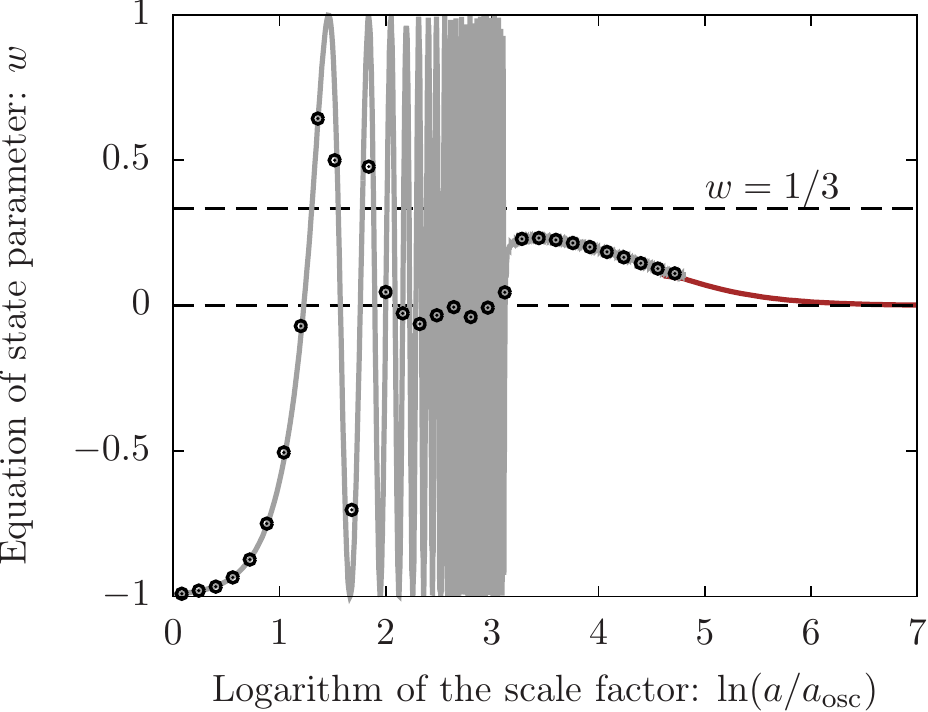}
\caption{Equation of state parameter $w$ as a function of the logarithm of the scale factor $\ln(a/a_{\mathrm{osc}})$. The continuous gray line is extracted numerically, from the linear mode evolution glued with the lattice simulation. The black dots represent the values of $\bar w$, averaged over short periods of time. The brown line is obtained by extrapolation in the frozen regime. We employ $\phi_1/f=1000$, $\kappa=10$ and $H_I/f=10^{-10}$.}
\label{eosDM}
\end{figure}

Our numerical simulations verify the above described behavior. As an illustration, in Fig.~\ref{eosDM} the equation of state parameter $w$ is shown as a function of $\ln(a/a_{\mathrm{osc}})$ for the case of $\phi_1/f=1000$, $\kappa=10$ and $H_I/f=10^{-10}$. The gray continuous line represents the numerically extracted $w$ (linear mode evolution glued with the lattice simulation). The black dots represent the values of $\bar w$, averaged over short periods of time. As expected, it shifts down as the background field oscillates with a smaller amplitude and probes the tachyonic regions of the potential. The oscillatory behavior of $w$ ends with the collapse of the background field.

As was already mentioned, the redshift and the finite lattice size restrict numerical simulations to not too late times. To access $w$ at late times, we performed the following extrapolation. 
As discussed in the previous subsection, at late times $a/a_{\rm osc}\gtrsim 100$ the Hubble expansion has sufficiently diluted the particles such that the rate of the self-interactions becomes small and the occupation numbers evolve very slowly. 
This allows us to approximate them as constant in comoving coordinates (as we have already done in the previous subsection). 

At late times, when the interaction rates drop, the pressure and the energy can be expressed in terms of the occupation numbers, 
\beq
\label{eq:pressure_energy}
\rho \approx  \frac{1}{a^4} \int_{\mathbf{p}} \omega(t,\mathrm{p}) n(t,\mathrm{p}), \: \: \: \: \: p \approx  \frac{1}{a^4} \int_{\mathbf{p}} \frac{\mathrm{p}^2}{3\omega(t,\mathrm{p})} n(t,\mathrm{p}).
\eeq
Here, $\omega^2(t, \mathrm{p}) \approx \mathrm{p}^2+a(t)^2m_a^2$ and the factors of $a^{-4}$ arise from the transformation from conformal variables back to the original ones. 

At late times the change of $w$ in time arises mostly from the growth of the scale factor, which enters the dispersion relation. We performed an extrapolation of $w$ to late times using (\ref{eq:pressure_energy}) with the latest available momentum distribution function. The result is represented by the brown line in Fig.~\ref{eosDM}. 

Having $w$ sufficiently close to zero at matter-radiation equality amounts essentially to the same constraint as discussed in the previous subsection.

\subsection{Energy density today}
\label{ssec:energy_density} 
The deviations from $w=0$ before the epoch of matter-radiation equality, which were discussed in the previous section, lead to an energy density today that is different from the one in the case of standard ALPs. 
While standard ALPs dilute as matter ($\bar w\approx 0$), already shortly after the oscillations start~\cite{Arias:2012az}, our monodromy ALPs have a significant phase
where $w\neq 0$. We now estimate the change in the dark matter density resulting from this.

Using the continuity equation
\beq
d[a^3(\rho+p)] = a^3dp
\eeq
and solving it for an arbitrary time-dependent or, equivalently, $a$-dependent parameter $w$ leads to
\beq
\rho(a)=\rho' \Bigl(  \frac{a}{a'}  \Bigr)^{-3} \exp \Big[-3\int_{a'}^a w(\widetilde a) d\ln \Bigl(  \frac{\widetilde a}{a'}  \Bigr) \Bigr].
\label{eq:dilution}
\eeq
Here $a'$ and $\rho'$ are the scale factor and the energy density at some reference time. In the case of a constant $w$ this simplifies to $\rho(a) \propto (a/a')^{-3(1+w)}$.

For monodromy ALPs the energy density today from (\ref{eq:dilution}) can be written as
\beq
\label{eq:Z}
\rho_{\mathrm{today}} = \rho_{\mathrm{osc}} \Bigl(  \frac{a_{\mathrm{today}}}{a_{\mathrm{osc}}}  \Bigr)^{-3} \mathcal{Z}_{\rho}(a_{\mathrm{today}}) ,\: \: \: \: \: \: \: \: \: \: \mathcal{Z}_{\rho}(a) = \exp \Big[-3\int_{a_{\mathrm{osc}}}^a w(\widetilde a) d\ln \Bigl(  \frac{\widetilde a}{a_{\mathrm{osc}}}  \Bigr) \Bigr].
\eeq
The dimensionless factor $\mathcal{Z}_{\rho}(a)$ shows the ratio of energy density at a scale factor $a$ to that in the case of pure matter-like dilution.

As it was already mentioned, if the background field ends up in a false minimum, some of its energy is transformed into the form of dark energy~\cite{Kobayashi:2018nzh}. This energy has to be additionally subtracted in the calculation of the dark matter density. 
Getting trapped in a false minimum is likely for large values of $\kappa$ and small values of $\phi_1/f$, such that\footnote{All local minima of the potential are in the range $-\kappa^2 \leq \phi/f \leq \kappa^2$.} $\phi_1/f \lesssim \kappa^2$. In this case the field typically gets trapped during the first couple of oscillations, or can also be trapped from the very beginning. The subsequent dynamics is then analogous to the case of standard ALPs and, in particular, the growth of fluctuations is insignificant.
In contrast, when $\phi_1/f$ exceeds $\kappa^2$, the field typically completes its oscillations around the lowest minimum. This is because the fluctuations manage to become strong and have the effect of smoothening out the wiggles in the effective potential, making it unlikely for the field to get trapped. 
In the following, we concentrate on the case of ending up in the true minimum, so that no dark energy contribution has to be subtracted. We return to the aspects of trapping in Section~\ref{ssec:power_spectrum}.

Excluding the possibility of ending up in a false minimum, 
$w$ inevitably approaches zero at sufficiently late times. In other words, $\mathcal{Z}_{\rho}$ becomes constant at large scale factors. This means that $\mathcal{Z}_{\rho}(a_{\mathrm{today}})$ in (\ref{eq:Z}) can be extracted from the simulations as the late-time nearly constant value of $\mathcal{Z}_{\rho}$, which we denote by $\mathcal{Z}_{\rho}^{\infty}$. For the energy density $\rho_{\mathrm{osc}}$ one can write
\beq
\rho_{\mathrm{osc}} \approx \frac{1}{2}m^2\phi_1^2 + \Lambda^4\Bigl[1-\cos\Bigl(\frac{\phi_1}{f}\Bigr)\Bigr] = \frac{1}{2}m_a^2 \phi_1^2 \mathcal{R}_{\kappa}\Bigl(\frac{\phi_1}{f}\Bigr),
\eeq
where
\beq
\mathcal{R}_{\kappa}(x) =  \frac{1}{1+\kappa^2} \Bigl(1+ \frac{2\kappa^2}{x^2}(1-\cos x) \Bigr).
\eeq
In order to simplify the expression for $\rho_{\mathrm{osc}}$, we focus on the case when $\phi_1/f$ significantly exceeds $\kappa$. This allows one to neglect the second term in $\mathcal{R}_{\kappa}$ and $\rho_{\mathrm{today}}$ from (\ref{eq:Z}) takes the form
\beq
\label{eq:energy_later}
\rho_{\mathrm{today}} =\Bigl[ \frac{1}{2}m_a^2 \phi_1^2 \Bigl(  \frac{a_{\mathrm{osc}} }{a_{\mathrm{today}}}  \Bigr)^{3} \Bigr] \Bigl( \frac{\mathcal{Z}_{\rho}^{\infty}}{1+\kappa^2} \Bigr).
\eeq
The expression in the squared brackets in the last equation is exactly what is calculated for the energy density of standard ALP dark matter today~\cite{Arias:2012az} and leads to (\ref{eq:dmdensitytd}). For monodromy ALP dark matter the expression for the energy density today is therefore modified to
\beq
\label{energy_today}
\rho_{\mathrm{today}} \approx 0.17\, \frac{\mathrm{keV}}{\mathrm{cm}^3} \sqrt{\frac{m_a}{\mathrm{eV}}} \Bigl(\frac{\phi_1}{10^{11}\mathrm{GeV}}\Bigr)^2 \Bigl( \frac{\mathcal{Z}_{\rho}^{\infty}}{1+\kappa^2} \Bigr).
\eeq

In the above expression, $\rho_{\mathrm{today}}$ is expressed in terms of the ALP mass $m_a$, the misalignment field value $\phi_1$, the parameter $\kappa$ and the factor $\mathcal{Z}_{\rho}^{\infty}$, which in turn depends on $\kappa$, $\phi_1/f$ and $H_I/f$. 

\begin{figure}[!t]
\centering
\includegraphics[width=0.63\textwidth]{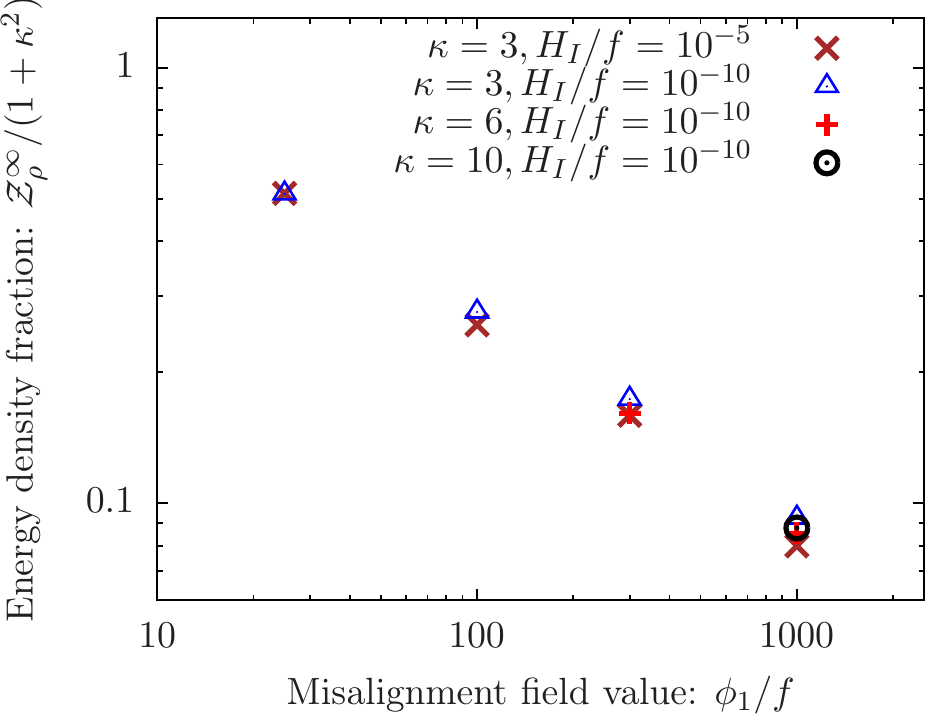}
\caption{The ratio of energy densities today of monodromy ALP dark matter and standard ALP dark matter, $\mathcal{Z}_{\rho}^{\infty}/(1+\kappa^2)$, as a function of the misalignment field value $\phi_1/f$, for different values of $\kappa$ and $H_I/f$.}
\label{Z}
\end{figure}

The values of $\mathcal{Z}_{\rho}^{\infty}/(1+\kappa^2)$, extracted from the simulations for different values of the parameters, are shown in Fig.~\ref{Z}. As can be seen in the plot, $\mathcal{Z}_{\rho}^{\infty}$ decreases with an increasing initial field value $\phi_1/f$, which is expected since more fluctuations are produced in that case, shifting $w$ more towards $1/3$. Remarkably, the dependence on the values of $H_I/f$ and $\kappa$ is very weak.

Overall, Fig.~\ref{Z} shows that the dilution due to the fluctuations is rather modest. The increased field amplitude easily compensates for this effect on todays energy density and the range of decay constants $f$ that allow for sufficient dark matter production can be extended in the presence of a monodromy. For example, at a fixed mass $m_{a}$ increasing the initial field value from $\field_{1}=f$ to $\field_{1}=1000\,f$ leads to an increase in today's energy density by a factor of $\sim 10^6\,Z^{\infty}_{\rho}/(1+\kappa^2)\gtrsim 10^5$. Stated differently, we can choose a value of $f$ that is smaller by a factor of $\sqrt{10^5}$ and have a correspondingly larger value of the coupling. This potential increase in the coupling is illustrated in Fig.~\ref{allowed} by the orange region, labelled ``Monodromy region".

\section{Strong fluctuations at small scales and their evolution}
\label{ssec:minicluster}

\subsection{Power spectrum of density fluctuations}
\label{ssec:power_spectrum}
As we have seen in Section~\ref{ssec:dm_growth}, for sufficiently large $\kappa$ and $\phi_{1}/f$ the energy density of the initially homogeneous background field is completely converted into fluctuations.
In that sense all of the dark matter is made from these (large) fluctuations.
To better understand the macroscopic picture it is useful to translate these field fluctuations into fluctuations in the energy density.

The energy density contrast is defined as
\beq
\delta(\mathbf{x})=\frac{\rho(\mathbf{x})-\langle \rho \rangle}{\langle \rho \rangle}.
\label{eq:ddc}
\eeq
With this the dimensionless density contrast power spectrum can be defined analogous to the field power spectrum, from the Fourier transform of the density contrast,
\beq
\label{eq:edps}
\langle \delta(t, \mathbf{p}) \delta(t, \mathbf{p'}) \rangle = (2\pi)^3\delta^{(3)} (\mathbf{p+p'})\frac{2\pi^2}{\mathrm{p}^3} \Delta_{\delta}(t, \mathrm{p}) .
\eeq
It is straightforward to check that the mean of the density contrast vanishes, 
\begin{equation}
\langle \delta(\mathbf{x})\rangle = 0,
\end{equation} 
while its variance can be expressed as 
\begin{equation}
\langle \delta^2(\mathbf{x})\rangle = \int d(\ln{\mathrm{p}}) \Delta_{\delta}(\mathrm{p}).
\end{equation} 

\bigskip

As already indicated, at early times energy is stored dominantly in the background field and the energy density contrast is very small, $\delta \ll 1$. After the fluctuations grow and the background field collapses, the power spectrum $\Delta_{\delta}$ becomes $\mathcal{O}(1)$\footnote{While these fluctuations are isocurvature in nature, their length scale is way too small to be constrained by the CMB measurements (cf.~\cite{Graham:2015rva,Alonso-Alvarez:2018tus} for similar situations).}. At late times the field dynamics slows down and, as a consequence, the power spectrum does not change with time anymore, where the overall red-shift due to expansion is included since comoving momenta are used. This holds until gravity becomes important, as we will discuss in the next subsection. 

\begin{figure}[!t]
\centering
\includegraphics[width=0.6\textwidth]{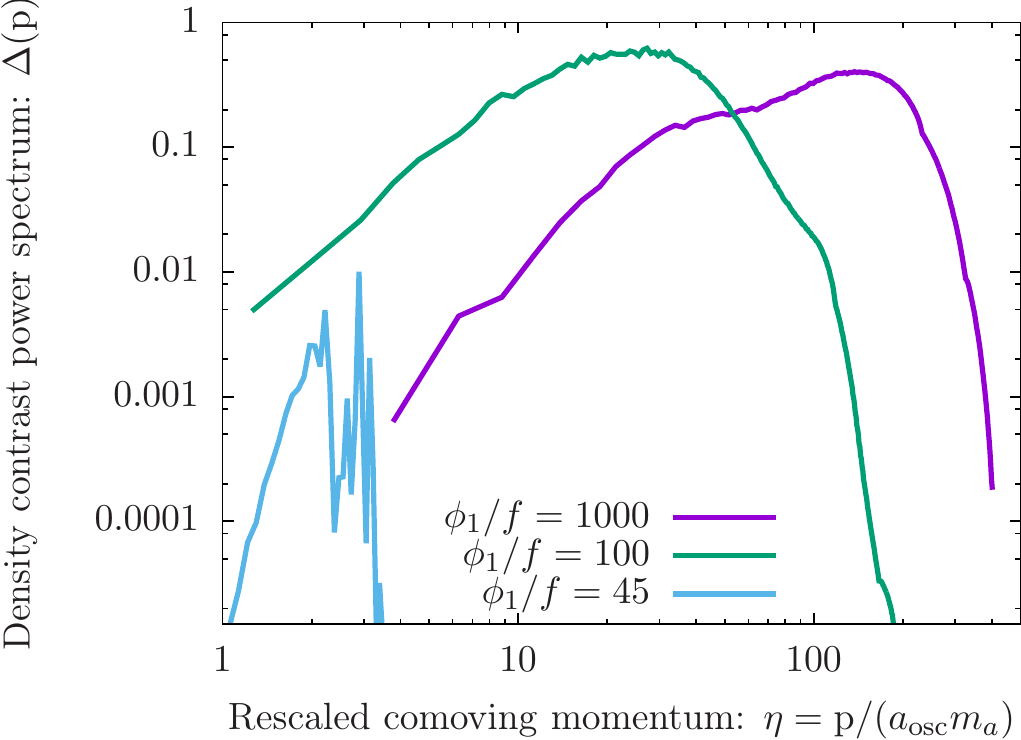}
\hspace{0.63cm}
\includegraphics[width=0.343\textwidth]{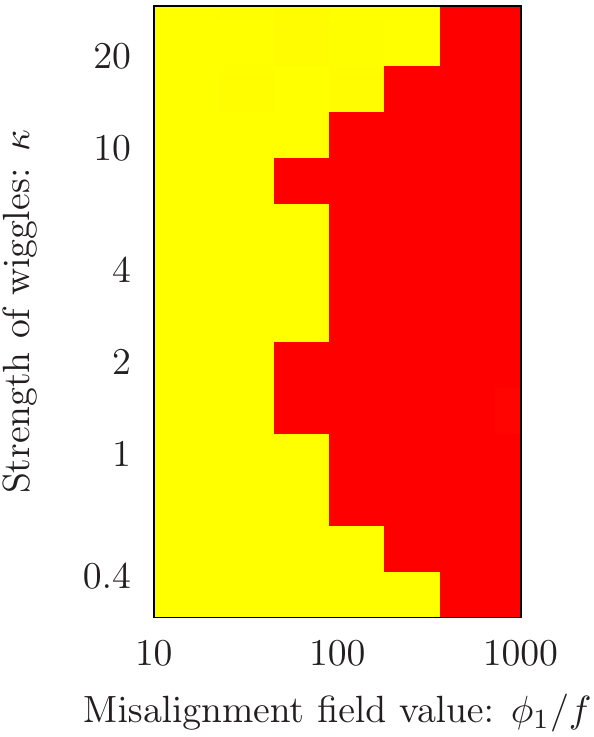}
\caption{Left: Several snapshots of the power spectrum, $\Delta(\mathrm{p})  =\mathrm{p}^3 \langle |\delta(\mathbf{p})|^2 \rangle/ (2\pi^2 V)$, as a function of the rescaled comoving momentum $\eta=\mathrm{p}/(m_a a_{\rm osc})$, for $\kappa=3$. Right: Fraction $r$ of the total energy density in the fluctuations. We have sampled $112$ different combinations of $\phi_1/f$ and $\kappa$. Points for which $r\ll1$ are marked in yellow, and those for which $r\approx 1$ are shown in red.  We employ $H_I/f=10^{-10}$.}
\label{PS}
\end{figure}

We have extracted the power spectrum $\Delta_{\delta}(t, \mathrm{p})$ numerically, from classical-statistical simulations\footnote{Alternatively one could extract it from the spectrum of field fluctuations as done, e.g. in~\cite{Graham:2015rva,Alonso-Alvarez:2018tus}.}. In the left panel of Fig.~\ref{PS} the late-time (frozen) power spectra are shown versus the comoving momenta, for two different values of the parameters for which the fluctuations become non-perturbatively large before freezing (green and purple). As can be seen in the plot, the power spectra have similar shapes. Remarkably, at their maximum $\Delta_{\delta} \approx 0.5$ in all cases, indicating that we indeed have large fluctuations in the density at relatively small scales i.e.~at $\eta \gg 1$, which is reminiscent of an inhomogeneous foam.
The locations of the maxima depend strongly on the initial field amplitude and very weakly on $\kappa$ and $H_I/f$.  While increasing $\phi_1/f$ pushes the location of the maximum towards higher momenta, the opposite does not work for small values of $\phi_{1}/f$. Indeed, if $\phi_1/f \sim10$ fluctuations remain small leading to a much weaker power spectrum. As an example, in the left panel of Fig.~\ref{PS} we show in blue the power spectrum for such a smaller initial field value, where fluctuations remain $\ll 1$.

Let us now discuss in more detail the parameters for which the over-densities become $\mathcal{O}(1)$. To illustrate this, in the right panel of Fig.~\ref{PS} we have plotted the fraction $r$ of the total energy density in the fluctuations\footnote{In (\ref{eq:r}), as well as in (\ref{eq:ddc}) we subtract the dark energy density, if its non-zero.}, 
\beq
r = \frac{\rho_{\rm fluc}}{\langle \rho \rangle}, 
\label{eq:r}
\eeq
for different values of $\phi_1/f$ and $\kappa$. The values of $r$ were obtained by performing numerical simulations with $112$ different combinations of $\phi_1/f$ and $\kappa$. We have set $H_I/f$ to $10^{-10}$, although the dependence on this parameter is rather weak. The yellow color in the plot corresponds to $r \ll 1$ i.e.~most of the energy being stored in the background field even at late times. The red color corresponds to $r\approx 1$. We have stopped the simulations once the energies in the oscillating background field and in the fluctuations became equal, since this is inevitably followed by the collapse of the background field. In other words, the red region in the plot corresponds to having $\mathcal{O}(1)$ over-densities at late times.

As can be seen, for $\mathcal{O}(1)$ values of $\kappa$, the over-densities become $\mathcal{O}(1)$ already for $\phi_1/f \sim 100$. For smaller values of $\kappa$ this transition occurs at larger field values, which is expected since the growth of fluctuations is slower in that case. Also in the upper part of the plot the transition to $r\approx 1$ occurs at larger field values. The reason is the high probability of getting trapped when $\phi_1/f\lesssim \kappa^2$, as it was explained in Section \ref{ssec:energy_density}. This prevents the fluctuations from growing significantly. Note, that even in the red region, in few cases the field ends up in a false minimum. Besides the reduction of the final dark matter density, this does not lead to any qualitative differences in the dynamics and, in particular, the power spectra have shapes similar to those in the left part of Fig.~\ref{PS}.

\subsection{No miniclusters}
The ${\mathcal{O}}(1)$ density fluctuations are reminiscent of the post-inflation axion scenario and it is tempting to consider
the formation of miniclusters as in~\cite{Hogan:1988mp,Kolb:1994fi,Enander:2017ogx,Fairbairn:2017sil}\footnote{Similarly for the case of dark matter created from inflationary ALP fluctuations~\cite{Alonso-Alvarez:2018tus}.}.

As shown in~\cite{Kolb:1994fi, Enander:2017ogx}, for pressureless matter the collapse of an over-density with a density contrast $\delta$ occurs at a scale factor,
\beq
\label{eq:crit_od}
x \approx \frac{0.7}{\delta}.
\eeq
where
\begin{equation}
\label{eqx}
a=x a_\mathrm{eq}, 
\end{equation}
and $a_\mathrm{eq}$ is the scale factor at the epoch of matter-radiation equality.

This indicates that ${\mathcal{O}}(1)$ fluctuations should collapse around the time of matter-radiation equality. However, in the case of monodromy dark matter we need to be a bit more careful.
The reason is that the length scale of the ${\mathcal{O}}(1)$ fluctuations is significantly smaller as can be seen from the fact that the peak of the power spectrum occurs at comoving momenta of the order (cf. Fig.~\ref{PS})
\begin{equation}
{\rm{p}}_{\rm max}/a_{\rm osc}\gtrsim 10 \,m_{a},
\end{equation}
instead of the $\lesssim m_{a}$ expected for a QCD axion in the post-inflation scenario\footnote{The QCD axion field has typically fluctuations of the size of the horizon at the time when the field starts to oscillate,
hence $\mathrm{p}\sim H_{\rm osc}\sim m(t_{osc})$. However, at this time the axion mass typically still increases due to the temperature dependence of the QCD effects giving rise to the potential. Therefore, the momentum measured in units of the final axion mass is actually smaller than the mass.}.

The length scale of the fluctuations is therefore significantly smaller. In terms of the particle language this simply means that the particles have a slightly higher velocity. As we will see, the corresponding small but non-vanishing pressure can have significant effects on the evolution.
To study this we use an adapted version of the Press-Schechter spherical collapse model used in~\cite{Kolb:1994fi, Enander:2017ogx} that includes these pressure effects in a similar manner to~\cite{Armitage:2007gn}. Following~\cite{Kolb:1994fi, Enander:2017ogx}, we first introduce the dimensionless variable $R$, which characterizes the deviation of the over-density from the background Hubble expansion, 
\beq
R(t)=\frac{r(t)}{r_{\rm flow}(t)} = \frac{r(t)}{a(t) \xi},
\eeq
such that $R=1$ before the over-density collapses and $\xi$ is the value of the comoving radius of the over-density at those early times. As shown in the appendix, in terms of $R$ and the variable $x$ from (\ref{eqx}) the evolution of the over-density is governed by
\beq
\label{final}
x\Bigl(1+x\Bigr)\ddot R  +  \Bigl(1+\frac{3}{2}x\Bigr)\dot R + \frac{1}{2}\Bigl[ \frac{1+\delta}{R^2}- R \Bigr] -   \frac{C}{xR^3} =0.
\eeq
Here the dots indicate derivatives with respect to $x$. The last term in (\ref{final}) is an extra-term due to non-vanishing pressure. 
Indeed, the only dependence on the size $\xi$ of the initial over-density is contained in the pressure term, or more precisely its pre-factor $C$. The pre-factor is calculated in the appendix and is given by
\beq
C= \frac{30}{\alpha^2}  \eta^4 \approx 0.76\, \eta^4 .
\eeq
Here $\eta = \mathrm{p}/(m_a a_{\mathrm{osc}})$ is the initial comoving momentum scale as in Fig.~\ref{PS}, and $\alpha\sim 2\pi$ is a factor that accounts for the dependence of the characteristic momentum scale of an over-density on its radius, i.e.~$\xi \sim \alpha/\mathrm{p}$. In the last step we set $\alpha = 2\pi$. 

Now we can find the condition for which the over-density can overcome the pressure force and collapse. Essentially, this requires that the gravitational attraction is stronger than the pressure, i.e.
\beq
\frac{\delta}{2R^2} > \frac{C}{xR^3}.
\eeq
This is equivalent to
\beq
x>x_C = \frac{2C}{\delta} = \frac{1.5 \eta^4 }{ \delta}.
\eeq

For fluctuations with $\delta\sim 1$ and  $\eta\sim {\rm few}\times10-100$ as expected from Fig.~\ref{PS}, we conclude that typical fluctuations have not yet collapsed~\cite{Fairbairn:2017sil}  today at $x_{\rm today}\sim 3000$. Therefore, we have strong fluctuations but not the huge over-densities $\gtrsim 10^9$ that we would expect for miniclusters. This is consistent with what one would expect from a Jeans analysis similar to the one in~\cite{Fairbairn:2017sil} i.e.~fluctuations below the Jeans scale do not grow. The difference in our case is that the fluctuations are already non-linear when they enter the matter-dominated regime. It is not clear whether non-linear effects in the presence of gravity and non-zero velocities can lead to a significant damping of these fluctuations. In any case we expect such a damping to be relatively slow.

\bigskip

Finally let us briefly attempt a first discussion of today's size of the fluctuations and the fate of the fluctuations during structure formation. 
The physical size of the fluctuation at a scale factor $a=x a_{\rm eq}$ is given by
\beq
\label{eq:typical_rad}
R  \sim \frac{a}{\rm p} \sim \Bigl( \frac{H_{\mathrm{osc}}}{H_{\mathrm{eq}}} \Bigr)^{1/2} \frac{1}{m_a \eta } x\sim   10^{4} \mathrm{km} \sqrt{ \frac{\mathrm{eV}}{m_a}} \frac{1}{\eta} x.
\eeq
If gravity plays no role we would expect for $\eta\sim {\rm few}\times 10-100$ typical sizes of the fluctuations of the order
\begin{equation}
R_{\mathrm{today}}\sim   (10^{5}-10^{6}) \mathrm{km} \sqrt{ \frac{\mathrm{eV}}{m_a}},
\end{equation}
where we have used  $x_{\rm today}\sim 3000$.

However, such small scale structures are part of the large scale structures such as galaxies and clusters. As they then only feel their local gravitational field, we expect that 
their Hubble growth stops as soon as the structure they are contained in decouples from the Hubble expansion.
Inside structures we therefore expect a somewhat smaller size of the fluctuations,
\begin{equation}
R^{\rm in \,\,structure}_{\mathrm{today}}\sim   (10^{5}-10^{6}) \mathrm{km} \sqrt{ \frac{\mathrm{eV}}{m_a}}\frac{a_{\rm structure}}{a_{\rm today}},
\end{equation}
where $a_{\rm structure}$ is the scale factor at the time when the structure decouples and therefore depends on the type of structure the fluctuation is contained within.
This can be as low as $10^3\,{\rm km}$ for early structures with $a_{\rm structure}/a_{\rm today}\sim 100$.

\section{Conclusions}
\label{ssec:conclusion}
Extending the field range  of axion-like particles (ALPs) by a monodromy provides a convenient way to increase the dark matter production, thereby allowing such ALPs
to be all of dark matter in a significantly enlarged parameter space. Here, potential couplings to Standard Model particles are larger and can be probed in near future experiments as shown in Fig.~\ref{allowed}.
The typical monodromy potential consists of a periodic part featuring a discrete shift symmetry as well as an explicit shift symmetry breaking quadratic  potential.
If the periodic potential is sizeable the potential exhibits pronounced wiggles that can lead to a significant and phenomenologically interesting change in the dynamics: it becomes dominated by fluctuations. 

In this paper we have used classical-statistical field theory simulations to track the evolution of these large, non-linear fluctuations through their whole evolution. Let us briefly summarize their evolution as well as the resulting potentially observable consequences.

As soon as the field begins to evolve, i.e.~at $H\lesssim m_a/3$, a rapid growth of fluctuations occurs. The fluctuations themselves quickly become non-linear dominating the energy density, whereas the homogeneous field component vanishes. The dynamics of such dark matter is therefore dominated by these fluctuations. This occurs well inside the radiation dominated regime and therefore much  earlier than in standard ALP DM, where the field remains essentially homogeneous until structure formation becomes important deep inside the matter dominated regime.
Indeed in a particle picture this growth of fluctuations corresponds to a conversion of the homogeneous background field into relativistic particles.  We find that physical momenta up to a ${\rm few}\times 100 \,m_a$ are significantly populated. Shortly after the field starts oscillating we therefore have a gas of relativistic particles and the equation of state is close to $1/3$, seemingly unsuitable for dark matter. However, as already mentioned, this occurs deep inside the radiation dominated area. The expansion effectively cools the gas 
and the equation of state returns to a value close to zero, as appropriate for cold dark matter. The relativistic period leads to a reduction in the total energy density compared to a situation where there is no significant growth of fluctuations and the equation of state remains zero during the whole evolution. While non-negligible, this effect does not destroy the possible enhancement in the density made possible by the enlarged field range. For example, for a field range enlarged by a factor of 1000 we obtain a naive enhancement of the dark matter density by a factor of $10^6$. For a typical situation with pronounced wiggles this is reduced by a factor $\sim 10$ due to the relativistic evolution, leaving most of the enhancement intact.

The most dramatic change from standard ALP dark matter lies in the structure at very small scales. Despite being relativistic (at very early times), the relevant momentum modes are very highly occupied. In terms of the density fluctuations this means that they are ${\mathcal{O}}(1)$. 
However, in contrast to the post-inflation axion scenario these structures do not collapse into miniclusters as their physical size is significantly smaller and the pressure from the resulting non-vanishing ALP velocity counteracts the gravitational force. Instead we have a ``foamy'' structure of the density at small scales with typical fluctuations of order one. This is probably hard to probe in astrophysical observations. However, if this structure survives inside the galaxy, it could lead to an  interesting fluctuating signal in Earth bound direct detection experiments (see e.g.~\cite{Horns:2012jf,Budker:2013hfa,Jaeckel:2013sqa,Chung:2016ysi,Graham:2015ifn,Kahn:2016aff,TheMADMAXWorkingGroup:2016hpc,Alesini:2017ifp,Melcon:2018dba,Du:2018uak,Carney:2019pza, Stadnik:2014tta}).

\section{Acknowledgements}
We thank Gonzalo Alonso-\'Alvarez, Francesco Cefala, Daniel Green, Doddy Marsh, Viraf Mehta and Bj\"orn Malte Sch\"afer for helpful discussions. 
A.C.~and J.J.~are grateful for support from the DFG via the transregional research collaborative TR33 “The Dark Universe”.
Some of the numerical calculations were performed on the computational resource bwUniCluster, funded by the Ministry of Science, Research and the Arts Baden-W\"urttemberg and the Universities of the State of Baden-W\"urttemberg, within the framework program Baden-W\"urttemberg high performance computing (bwHPC).

\appendix

\section{Appendix: The modified spherical collapse model}
\label{app}

In this appendix we describe the modified version of the spherical collapse model~\cite{Kolb:1994fi, Enander:2017ogx}, which takes into account the non-vanishing pressure of the over-densities of monodromy ALP dark matter.

We consider an over-density of radius $r$, density $\rho$, pressure $p$ and total mass $M$. The Newtonian equation of motion for the radius has the form
\beq
\label{newt}
\frac{d^2r}{dt^2} = -\frac{8\pi G}{3} \rho_R r -\frac{GM}{r^2}-\frac{1}{\rho} \frac{dp}{dr},
\eeq
where $G$ is the Newton's constant and $\rho_R$ is the background density of radiation. The last term is the acceleration due to the pressure~\cite{Armitage:2007gn}. The total mass is conserved and related to the initial\footnote{The density contrast starts to grow after the over-density collapses.} density contrast $\delta$ of the over-density by
\beq
M=\frac{4\pi}{3}\rho r^3 = \frac{4\pi}{3}\bar \rho  (1+\delta) r^3.
\eeq

The pressure term can be estimated by noting that ALPs have already become non-relativistic by the time of interest, such that $\epsilon_{\mathbf{p}}\approx m_a$. Therefore,
\beq
\rho \approx m_a \frac{N}{V}=\frac{M}{V}\propto r^{-3},  \: \: \: \: \: \: \: \: \: \: \: \: p  \approx \frac{\langle \mathrm{p}^2 \rangle}{3m_a} \frac{N}{V} \approx \frac{\langle \mathrm{p}^2 \rangle}{3m_a^2} \rho = \frac{\alpha^2}{3m_a^2 r^2} \rho \propto r^{-5}.
\eeq 
where $N$ is the number of ALPs in the over-density and $\alpha \sim 2\pi$ relates the radius to the characteristic momentum. The acceleration due to the pressure is then given by
\beq
-\frac{1}{\rho} \frac{dp}{dr} = \frac{1}{\rho} \frac{5p}{r}= \frac{5\alpha^2}{3m_a^2 r^3}.
\eeq

The radius of the over-density can be expressed as $r(t)=\xi a(t) R(t)$, where $\xi$ is the comoving radius before decoupling and $R(t)$ describes the deviation of the radius from the background flow Hubble flow (such that $R=1$ at early times, before decoupling). Transforming from time to the variable $x=a/a_{\rm eq}$, as explained in~\cite{Kolb:1994fi}. in our case leads to
\beq
\label{final:sphcol}
x\Bigl(1+x\Bigr)\ddot R  +  \Bigl(1+\frac{3}{2}x\Bigr)\dot R + \frac{1}{2}\Bigl[ \frac{1+\delta}{R^2}- R \Bigr] -   \frac{C}{xR^3} =0.
\eeq
The last term in (\ref{final:sphcol}) is the extra-term due to the non-vanishing pressure. The pre-factor $C$ is given by
\beq
C= \frac{5 \alpha^2}{8\pi G \bar \rho_{\mathrm{eq}}a_{\mathrm{eq}}^4 \xi^4 m_a^2},
\eeq
where $\bar \rho_{\mathrm{eq}}$ is the average density of ALP dark matter at the epoch of matter-radiation equality. 

The expression for $C$ can be found by relating the comoving radius of the over-density to its characteristic comoving momentum, $\xi = \alpha/\rm p$. Using (\ref{scale_factor_Hubble}), this leads to
\beq
C=\frac{5 \alpha^2}{8\pi G \bar \rho_{\mathrm{eq}}}\frac{p^4}{m_a^2 a_{\mathrm{eq}}^4 \alpha^4} = \frac{5 \alpha^2}{8\pi G \bar \rho_{\mathrm{eq}}} \frac{ 9 H_{\mathrm{eq}}^2 \eta^4}{\alpha^4},
\eeq
where $\eta = {\rm p}/(m_a a_{\mathrm{osc}})$. Finally, noting that at the epoch of matter-radiation equality $H_{\mathrm{eq}}^2 = (8\pi G/3) 2\bar \rho_{\mathrm{eq}}$, we arrive at
\beq
C(\eta) = \frac{30}{\alpha^2}  \eta^4 \approx 0.76\, \eta^4 ,
\eeq
where in the last step we set $\alpha \approx 2\pi$. As can be seen, the parameter $C$ is proportional to the fourth power of the typical dimensionless rescaled comoving momentum $\eta$ of the over-density.

\end{document}